\documentclass[12pt]{article}
\usepackage{latexsym,epsfig,amssymb,euscript,amsmath,verbatim}

 \def\be{\begin{equation}} \def\ee{\end{equation}}
\def\bea{\begin{eqnarray}} \def\eea{\end{eqnarray}}
\newcommand{\nn}{\nonumber} 
 
\newcommand{\non}{\nonumber}

\begin{document}

\pagestyle{empty}
\rightline{SISSA-71/2009/EP}

\rightline{ULB-TH/09-40}
\vspace{0.8cm}
\begin{center}
{\Large{\bf Patterns of Soft Masses from \\
\vspace{4mm}
General Semi-Direct Gauge Mediation}}

\vskip 20pt
 {\large{Riccardo Argurio$^{1}$,
Matteo Bertolini$^{2}$, Gabriele Ferretti$^{3}$ \vskip 3pt and  Alberto Mariotti$^{4}$ \\[5mm]}}

{\small{{}$^1$ Physique Th\'eorique et Math\'ematique and International Solvay
Institutes \\
\vspace*{-2pt}  Universit\'e Libre de Bruxelles, C.P. 231, 1050
Bruxelles, Belgium\\

\medskip
{}$^2$ SISSA and INFN - Sezione di Trieste\\
\vspace*{-2pt} Via Beirut 2; I 34014 Trieste, Italy\\

\medskip
{}$^3$  Department of Fundamental Physics \\
\vspace*{-2pt} Chalmers University of Technology, 412 96 G\"oteborg, Sweden
\\

\medskip
{}$^4$ Theoretische Natuurkunde and International Solvay Institutes \\
\vspace*{-2pt}
Vrije Universiteit Brussel, Pleinlaan 2, B-1050 Brussels, Belgium}}\\

\medskip

\medskip

\medskip

\medskip

{\bf Abstract}
\vskip 20pt
\begin{minipage}[h]{16.0cm}

We give a general formulation of semi-direct gauge mediation of supersymmetry
breaking where the messengers interact with the hidden sector only through a
weakly gauged group. Using this general formulation, we provide an explicit
proof that the MSSM gaugino masses are vanishing to leading order in the gauge
couplings. On the other hand, the MSSM sfermion masses have, generically, a
non-vanishing leading contribution. We discuss how such a mechanism can
successfully be combined with other mediation schemes which give tachyonic
sfermions, such as sequestered anomaly mediation and some direct gauge
mediation models.

\end{minipage}
\end{center}
\newpage
\setcounter{page}{1} \pagestyle{plain}
\renewcommand{\thefootnote}{\arabic{footnote}} \setcounter{footnote}{0}

\section{Introduction}

Gauge mediation \cite{original} (see \cite{Giudice:1998bp} for a comprehensive review) is a popular mechanism to
mediate supersymmetry breaking from a hidden sector, where the breaking occurs, to the visible sector (the MSSM or
extensions thereof). Its main virtue is, besides calculability, that of naturally implementing a strong
suppression of all soft terms leading to flavor changing neutral
currents. In this framework, the only degrees of freedom of the MSSM
which interact with the hidden sector are those of the gauge sector.

In exploring the phenomenological implications of various mechanisms of supersymmetry breaking, it is desirable to
isolate those predictions that are independent of the specific
details of the model. Recently, a general formulation of gauge
mediation (GGM) was given that accomplishes this in a very explicit way \cite{Meade:2008wd}. In this formalism,
gauge mediation is formulated in terms of the currents (both fermionic and bosonic) that couple to the gauge
degrees of freedom. It can be shown that the spectrum of soft masses is restricted by two sum rules for the
sfermions but is otherwise generic. For instance, these sum rules do not specify any
pattern between the masses of gauginos and the masses of the sfermions, which is thus an
undefined feature in a generic gauge mediation model. For a discussion
of such patterns and hierarchies, see e.g.~\cite{Buican:2008ws,Abel:2009ze}.

In practice, specific models of gauge mediation will prominently feature messenger superfields, which are
those fields charged under the MSSM gauge groups that also interact with the supersymmetry breaking sector.
Once a given model is specified, patterns and hierarchies among the soft terms immediately arise. As an example,
the minimal gauge mediation scenario (MGM) \cite{messengers} (often used in phenomenological applications)
has messenger chiral superfields which couple trilinearly with a spurion that provides them with both a supersymmetric
mass and an off-diagonal non supersymmetric mass. In this model the gaugino and sfermion masses turn out
to be of the same order of magnitude. Other examples, which are denoted direct gauge mediation
models (DGM) \cite{direct}, have messengers which are typically composite fields directly participating in the
dynamical supersymmetry breaking mechanism. Clearly,
those would be the most appealing models (solving the
hierarchy problem with no tuned parameters), but it turns out that
often in such models the gaugino masses are highly
suppressed or the sfermion masses are tachyonic. In addition, supersymmetry breaking generally requires a
large hidden gauge group giving rise to a Landau pole in the visible
couplings. Though models which cure some of these problems exist
(possibly based on metastable hidden sectors, see e.g.
\cite{koolike}), it would be desirable to single out general properties of models which are phenomenologically
viable.

We focus on a class of models where the messengers interact with the hidden sector only through
(non-MSSM) gauge interactions with gauge group $G_h$  and coupling $g_h$ but, 
unlike DGM, they do not participate to the supersymmetry breaking dynamics.
These models were dubbed semi-direct gauge mediation (SDGM) in~\cite{Seiberg:2008qj}.
A subclass of these models -- characterized by the further requirement that letting the hidden coupling $g_h \to 0$ did not lead to restoration of supersymmetry in the hidden sector -- was considered in~\cite{Randall:1996zi} under the name of mediator models.
Mediator models obey directly all the conditions of GGM \cite{Meade:2008wd} if one considers the messengers as part of the visible sector.
Our computations are done mainly with this class of models in mind. We believe, however, that our analysis can be extended even to the class of models of SDGM considered in \cite{Seiberg:2008qj} -- characterized by the fact that supersymmetry breaking requires  
$g_h \not = 0$ -- as long as one is allowed to treat perturbatively the masses of the hidden gauge multiplets, although a full analysis of the issues involved in higgsed gauged mediation is beyond the scope of the present paper.

SDGM models lie somewhat in between minimal and direct gauge mediation. Like MGM they have an explicit messenger sector. Like DGM,
however, no spurion-like coupling is needed and everything is mediated by
gauge interactions alone. The only superpotential term for
the messenger field  is a mass term. From a theoretical point of view the interest of such models lies in their simplicity,
and also in the rather straightforward way in which they can be generated in string theory inspired quiver gauge theories
(in which also the mass term arises dynamically, without the need to
be introduced as an external fine tuned parameter) \cite{Argurio:2009pz}.
An important advantage
of these models with respect to DGM is that they ameliorate the Landau pole problem which often afflicts DGM models, since
$G_h$ can be as small as $U(1)$.

Our goal is to discuss the generic features of this class of models by implementing a formalism very similar to the one
of GGM \cite{Meade:2008wd}. Our approach will be general since we parameterize the supersymmetry
breaking sector by currents instead of by a spurion. We aim at computing general expressions relating the MSSM
gaugino and sfermion masses to the correlators of the supersymmetry breaking currents.

The outcome of our analysis is as follows.

The gaugino masses are vanishing at the first order where they would be expected to appear. This agrees with the results obtained
both in \cite{ArkaniHamed:1998kj}
and in \cite{Seiberg:2008qj} (see also \cite{Ibe:2009bh}). In all of those papers, an effective approach was
used to provide the argument of vanishing gaugino masses. Here we provide a derivation of this result based on the precise cancellation
between the two diagrams contributing to the gaugino mass. This cancellation takes place for any supersymmetry breaking
current correlator. Hence, the cancellation is a result which does not depend on the existence of a hidden sector
spurion encoding supersymmetry breaking, nor on the specific
supersymmetry breaking mechanism occurring in the hidden sector. (Note that
this cancellation also invalidates the possibility for unsuppressed gaugino mass in one of the string-inspired quiver gauge
theory models discussed in the final section of~\cite{Argurio:2009pz}.) 

The sfermion masses on the other hand do not vanish at the first order in which they are expected to
appear, namely, forth order in both the hidden and the visible gauge coupling ($g_h^4 g_v^4$). We provide an expression for these masses which is very reminiscent
of the one appearing in general gauge mediation, though there is a complicated kernel appearing in
the momentum integral over the hidden sector correlators. This kernel has two effects. Firstly it reverses the sign
of the supertrace (i.e., if the hidden sector would have given tachyonic sfermion masses in a direct gauge mediation
scenario, it will give positive squared masses in a semi-direct gauge mediation scenario). Secondly, it has a soft
behavior at low momenta, and a mild logarithmic growth at large momenta, such that the sfermion masses are safely
under control and finite, even though we generically have a
non-vanishing supertrace in the messenger sector. This
is to be confronted with \cite{Poppitz:1996xw} where a messenger
supertrace was introduced by hand and led to UV
divergent (and hence UV sensitive) sfermion masses, due to the fact
that once some soft scalar masses are introduced, the others necessarily
undergo renormalization.

One might conclude that semi-direct gauge mediation, having almost vanishing
gaugino masses, is not phenomenologically interesting. On the contrary, we will argue that it can be quite
useful when combined with other mechanisms of mediation of supersymmetry breaking, either with a single or multiple hidden
sectors. In particular, one could think  of combining semi-direct gauge mediation with anomaly mediation (with a
sequestered hidden sector) \cite{Randall:1998uk}. As we shall see, the sfermion mass contribution from SDGM can stabilize
the otherwise tachyonic sleptons arising in the simplest models of anomaly mediation.
Interestingly, contrary to what one may superficially
imagine, it turns out that in this scenario no substantial fine-tuning is required to accomplish such a welcome
conspiracy between the two competing effects. We also discuss another scenario where one combines
SDGM with models of direct gauge mediation. As we will show in detail, at fixed hidden sector, SDGM and direct gauge
mediation provide opposite signs for the squared sfermion masses. Hence in this case SDGM can be useful both in combination
with models with tachyonic sfermions, as before, or in models with suppressed gauginos to make
the MSSM sparticle spectrum  all of the same order (or even to invert the hierarchy between sfermions and gauginos).
In both these cases, however, differently from anomaly mediation, generically one would need a fine tuning. For related work on conformal gauge mediation, see \cite{Ibe:2007wp}.

The rest of this paper is organized as follows. In section 2 we
present the semi-direct gauge mediation adapted version of the GGM
formalism and explain how to use it to compute the visible mass
spectrum. In section 3 we compute the gaugino masses and in
section 4 the sfermion masses. We end in section 5 discussing the
possible phenomenological relevance of semi-direct gauge mediation
suggested by our analysis.

\section{General set up for semi-direct gauge mediation}

\begin{figure}[ht]
\centering
\includegraphics[width=0.8\textwidth]{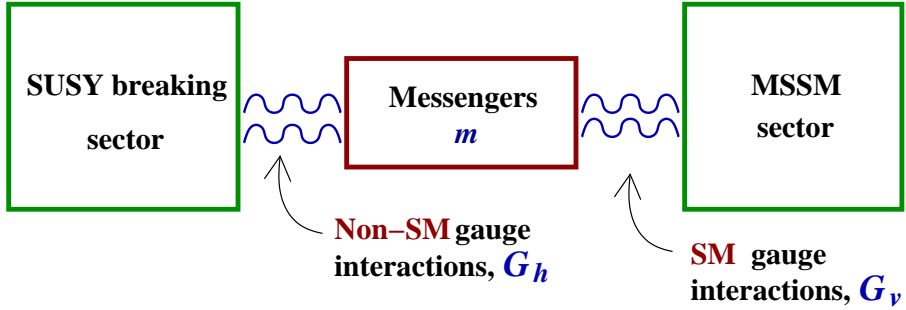}
\caption{\small A schematic picture of semi-direct gauge mediation. The gauge group $G_h$ is
singled-out within the hidden sector as the subgroup of the hidden gauge group to which the
messenger superfields couple. Messengers have a supersymmetric mass $m$.
\label{sdgm}}
\end{figure}

The models we are considering are characterized by three building blocks:
\begin{itemize}
\item{\vskip -5pt A visible sector with gauge group $G_v$ (the MSSM or any extension thereof).}
\item{\vskip -7pt A hidden supersymmetry breaking sector containing, besides confining gauge groups driving dynamical
supersymmetry breaking, a global continuous symmetry group $G_h$, which is then weakly gauged.}
\item{\vskip -7pt A pair of messenger superfields $\Phi$ and $\tilde \Phi$ in the bi-fundamentals of $G_v$ and $G_h$,
having a supersymmetric mass $m$ but no other superpotential interactions.}
\end{itemize}
\vskip -5pt In this section, for simplicity, we assume that both
$G_h$ and $G_v$ are $U(1)$ factors. All our results easily generalize to arbitrary
gauge groups by adding the appropriate group theory factors. A
pictorial representation of the SDGM scenario is reported in
figure \ref{sdgm}.

\vskip 6pt
\vskip -4pt

In the limit where the gauge coupling $g_h$ of the gauge group $G_h$ is sent to zero, the whole system separates into
two completely decoupled ones: the supersymmetry breaking sector and a supersymmetry preserving sector comprising the messenger
sector and the MSSM fields. This property allows us to use an approach similar to \cite{Meade:2008wd}, as we can parametrize all of the
supersymmetry breaking effects through the correlators of the global $G_h$ currents. Namely, we can write, in momentum
space\footnote{We use a slightly different sign convention with respect to \cite{Meade:2008wd}, see Appendix A.}
\bea
\langle J^h(p) J^h(-p)\rangle & = & C^h_0 (p^2/M^2)~, \nn \\
\langle j^h_\alpha (p) \bar j^h_{\dot{\alpha}}(-p)\rangle & = & p_\mu \sigma^\mu_{\alpha\dot{\alpha}}C^h_{1/2}
(p^2/M^2)~, \nn \\
\langle j^h_\alpha(p) j^h_\beta (-p)\rangle & = & \epsilon_{\alpha\beta} M B^h (p^2/M^2)~, \label{firstequation} \\
\langle j^h_\mu(p) j^h_\nu (-p)\rangle & = & (p_\mu p_\nu - p^2 \eta_{\mu\nu}) C^h_1 (p^2/M^2)~, \nn
\eea
where $J^h$, $j^h_\alpha$ and $j_\mu^h$ are respectively the scalar, spinor and vector components of the
current superfield ${\cal J}^h$, and $M$ is a typical scale of the dynamical supersymmetry breaking sector. If there
is more than one scale in that sector, the hierarchies will be encoded in small numerical constants in the functions
$C^h_s$ and $B^h$. Additionally, we could also have a non vanishing one-point function for $J^h$
\be
\langle J^h \rangle = D^h~.
\ee
This expectation value is identified with a non zero $D$-term in the gauge
group $G_h$, which is then spontaneously broken. Our approach can accommodate easily such a
situation, provided the group $G_h$ is weakly gauged and the corresponding Higgs generated masses can be treated perturbatively
(as in \cite{Buican:2009vv}). However we will actually find that such
$D^h$ will not contribute to the soft masses at leading order.

The complete Lagrangian of the model reads
\bea
\label{Lsm}
{\cal L} & = & {\cal L}_\mathrm{MSSM}  + \int d^4\theta \big( \Phi^\dagger e^{g_v V_v  + g_h V_h} \Phi
+\tilde \Phi e^{-g_v V_v  - g_h V_h} \tilde\Phi^\dagger \big) \nonumber \\
& & + \int d^2\theta \,m \Phi \tilde \Phi + \int d^2 \theta \,\mathrm{tr} {\cal W}_h^2  + h.c. +
\int d^4 \theta \, g_h V_h {\cal J}^h~, \label{Lsemi}
\eea
where, with obvious notation, $g_h$ is the gauge coupling of the gauge group
$G_h$, and $g_v$ that of the visible sector gauge group $G_v$. The last term in (\ref{Lsemi}) represents the first order
coupling to the supersymmetry breaking dynamics encoded in (\ref{firstequation}). There are, of course, additional non linear
terms required by   gauge invariance and supersymmetry.

In contrast, in general gauge mediation one writes for the same Lagrangian\footnote{The superscript $v$ in ${\cal J}^v$ refers
to the fact that the currents couple to the \emph{visible} gauge
fields.}
\be
{\cal L}  =  {\cal L}_\mathrm{MSSM}  + \int d^4 \theta \, g_v V_v {\cal J}^v~.\label{Ldir}
\ee
In turn one encodes all the mediation of supersymmetry breaking to the visible gauge sector in the functions
$C^v_s$ and $B^v$ characterizing the correlators of the various components of ${\cal J}^v$.
We thus see that we have in this class of models a sort of {\it tumbling} where the functions $C^v_s$ and $B^v$ are
eventually determined in terms of the functions $C^h_s$ and $B^h$. This will be our approach in
computing the soft masses of the MSSM.

It is rather straightforward to write the general structure of the Feynman diagrams allowing us to extract
$C^v_s$ and $B^v$ from $C^h_s$ and $B^h$. What we have to compute are radiative corrections to the two-point functions
of the fields in the visible gauge sector. The only relevant radiative corrections will be those involving messenger
fields, which in turn will be corrected by virtual particles of the gauge group $G_h$. In order to have
supersymmetry breaking at all, the propagators of the latter must include the insertion of a hidden current correlator.

\begin{figure}[ht]
\vskip 7pt
\centering
\includegraphics[width=1\textwidth]{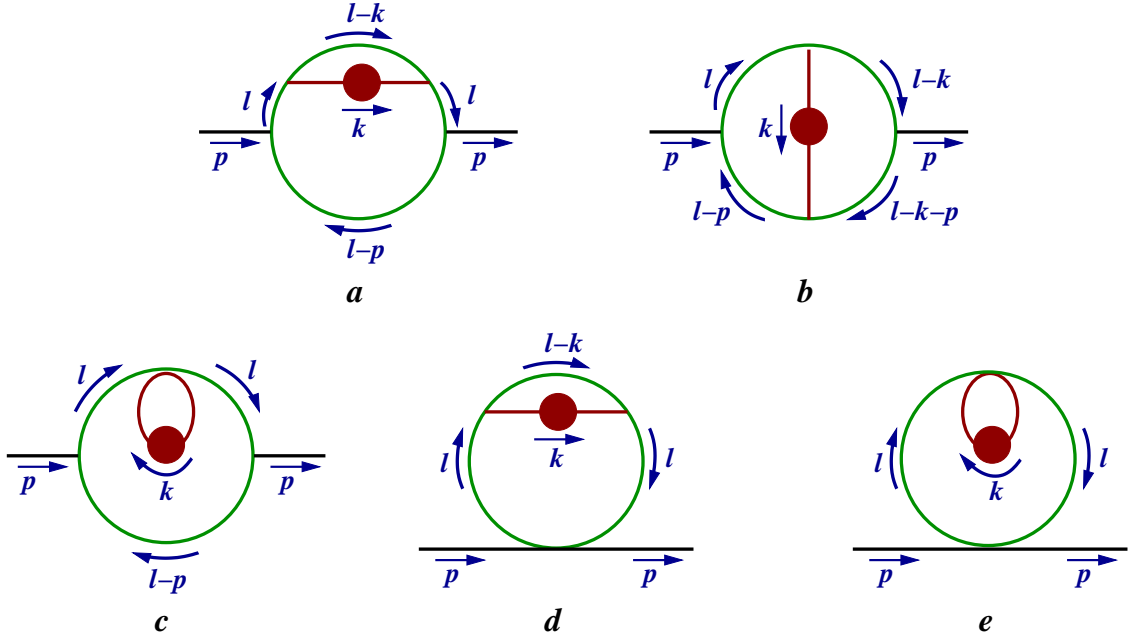}
\caption{\small The five topologically inequivalent structures contributing to the visible GGM
functions. External lines, which carry momentum $p$, correspond to $G_v$-fields ({\it i.e.} fields
belonging to the visible gauge group). Internal lines carrying momentum $k$ correspond to $G_h$-fields ({\it i.e.}
fields belonging to the hidden sector gauge group $G_h$) and have attached a blob which encodes the exact hidden
sector non-supersymmetric correction to the corresponding propagators. Finally, all other internal lines correspond
to messenger fields.
\label{top}}
\end{figure}

In figure \ref{top} we have drawn the five topologically distinct Feynman diagrams entering the computation.

The external lines are those of the $G_v$-fields (the MSSM gauge bosons, gauginos and
auxiliary $D$-fields). Internal lines carrying momentum $k$ are associated to the fields belonging to the
gauge group $G_h$ degrees of freedom, and hence will also involve a correlator insertion. All other
internal lines are messenger lines, either scalar or fermionic. All these diagrams have two explicit loops, plus
an additional loop factor coming from the hidden correlator insertion. Hence they all scale like $g_v^2 g_h^4$.

A first observation is that while the diagrams of topology $a$, $c$, $d$ and $e$ can be effectively taken into
account as one loop diagrams with corrected messenger lines (i.e. they involve only messenger two-point functions),
the diagram with topology $b$ is intrinsically two loops since it involves a messenger four-point function. The
contribution of these latter diagrams cannot be encoded in an effective approach such as the one of
\cite{Poppitz:1996xw}. As we will see, in computing the visible gaugino masses, there is a dramatic cancellation
between the two contributions from the diagrams $a$ and $b$.

A second important observation is that the correlator encoded by the
function $B$, which is complex, is on a different footing with
respect to the others, which are real. By carefully keeping track of the chiral structure of each diagram, one can see that $B^v$ is
a function of $B^h$ only, while the $C^v_s$ are functions of a linear combination of the $C^h_s$.

The knowledge of the functions $C^h_s$ and $B^h$ and, via
tumbling, of the functions $C^v_s$ and $B^v$, completely determines
the soft masses in the MSSM Lagrangian (\ref{Lsm}). In what follows, we will
compute, using the above formalism, both gaugino and sfermion
masses to leading order. Obviously, being a model of gauge mediation, general SDGM obeys, if considered in isolation
from other mechanisms, the same sfermion sum rules of GGM~\cite{Meade:2008wd}.

\section{Vanishing of gaugino masses}

In this section we compute the gaugino masses. For this computation, the diagrams of
figure \ref{top} are really the end of the story since the external lines are nothing but the gauginos themselves,
and we can set the external  momentum to be $p=0$.

\begin{figure}[ht]
\centering
\includegraphics[width=0.85\textwidth]{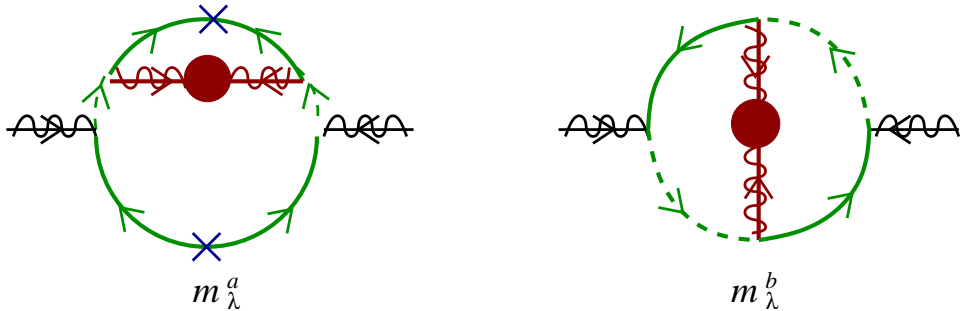}
\caption{\small The diagrams contributing to the gaugino mass. The
left one has two (supersymmetric) mass insertions, each one represented by a cross on the
corresponding messenger fermionic line.
\label{gaugino}}
\end{figure}

Of the five topologically inequivalent structures of figure
\ref{top} only two contribute to the gaugino masses. 
Indeed, diagrams $d$ and $e$ appear only if the external lines are vector 
bosons. Diagram $c$ appears only if the internal line (the one with the blob 
attached) is a vector bosons, and diagrams of this type are prevented 
from contributing to gaugino masses by chirality.
The only relevant ones are then those of type $a$ and
$b$. Their precise structure is depicted in figure \ref{gaugino}.

The first one is the same as the one discussed in
\cite{Randall:1996zi} and evaluates to\footnote{For convenience we drop the gauge theory factor common to both diagrams.
Our result is valid for any gauge group.}
\be
m^{a}_{\lambda} =
-16 \, g_v^2 \, g_h^4 \int \frac{d^4 k}{(2\pi)^4} \int \frac{d^4 l}{(2\pi)^4}
\frac{m^2 M B^h(k^2/M^2) }{k^2 \, (l^2+m^2)^3 \,[(l-k)^2+m^2]}~.
\ee
These are Wick rotated expressions and the conventions used for the Euclidean propagators are summarized in
Appendix A. We have factors of $4$ coming from the four Yukawa vertices, $2$ coming from the trace on the internal
fermion loop (which also gives the overall $-$ sign), and $2$ from the messenger multiplicity.

\vskip 10pt

The second diagram, of type $b$, gives the following contribution
\be
m^{b}_{\lambda}= 8 \, g_v^2 \, g_h^4 \int \frac{d^4 k}{(2\pi)^4} \int \frac{d^4 l}{(2\pi)^4}
\frac{l \cdot (l-k)  \, M B^h(k^2/M^2) }{ k^2 \, (l^2+m^2)^2 \, [(l-k)^2+m^2]^2
}~.
\ee
The factors are as before, except for the missing $-2$ since we have a continuous fermionic line instead of a
fermion loop. Obviously, we have that
\be
m_\lambda = m^{a}_{\lambda} + m^{b}_{\lambda}~.
\ee

In the two expressions above, the integral over the $l$-momentum can be
done analytically by standard techniques. We can then write
\be
m^{a,b}_{\lambda} = 8 \, g_v^2 \, g_h^4 \, \frac{M}{m^2} \int \frac{d^4 k}{(2\pi)^4} \frac{1}{k^2} L^{a,b} (k^2/m^2)
\, B^h(k^2/M^2)~,
\ee
where the two kernels are given by
\bea
L^a (k^2/m^2) & = &-2 \, m^4  \int \frac{d^4 l}{(2\pi)^4} \frac{1}{(l^2+m^2)^3 \, [(l-k)^2+m^2]}~, \\
L^b (k^2/m^2) & = &m^2  \int \frac{d^4 l}{(2\pi)^4} \frac{l\cdot (l-k)}{(l^2+m^2)^2 \, [(l-k)^2+m^2]^2}~.
\eea
Evaluating the two integrals above we find
\be
L^a(x) = - L^b(x) = -\frac{1}{32 \pi^2}\big( \frac{1}{x} + \frac{1}{x+4} - \frac{4f(x)}{x(x+4)}\big)~,
\ee
where\footnote{For the record, note that the same function $f(x)$ appears when one evaluates the function $B^v$ in
a minimal gauge mediation scenario $W=(m+\theta^2 F) \Phi \tilde \Phi$, namely $B^v(p^2) = (1/32 \pi^2) (F/m^2) f(p^2/m^2)$.}
\be
f(x) = \frac{4}{\sqrt{x(x+4)}}\, \mathrm{arctanh} \sqrt{\frac{x}{x+4}}
\ee
and $x=k^2/m^2$. This result means that the total kernel in the expression for the gaugino mass
\be
m_{\lambda} = 8\, g_v^2 \, g_h^4\, \frac{M}{m^2} \int \frac{d^4 k}{(2\pi)^4} \frac{1}{k^2} L (k^2/m^2)
\, B^h(k^2/M^2)
\ee
vanishes, since
\be
L(x) = L^a(x) + L^b(x) = 0 ~.
\ee
Hence, $m_\lambda=0$ for any function $B^h$, at this order. The fact that the gaugino mass was vanishing at leading order in
this class of models was first noted in \cite{ArkaniHamed:1998kj} where an effective argument
based on wave function renormalization was given. There it is also
shown that the visible gaugino first obtains a mass at order $g_h^4 g_v^6$. This effect has been dubbed gaugino screening.

Here we have re-derived this important result in a different way.
We have shown explicitly how the cancellation arises, which was not obvious from the start
(indeed, even a posteriori, the cancellation would seem rather miraculous if we did not have an independent
argument in favor of it taking place). Also, and perhaps more importantly, the proof that the kernel $L(x)$
is zero means that the cancellation does not depend on $B^h$, and hence applies to any model of supersymmetry
breaking that is responsible for the presence of $B^h$. This is a counterpart to the argument given in
\cite{ArkaniHamed:1998kj} which hinges on the capability of
effectively encoding the breaking of supersymmetry in a spurion field,
and on the assumption that there are hidden messengers mediating supersymmetry breaking to the
gauge group $G_h$.

\section{Sfermion masses}

We now turn to the computation of the general expression for the MSSM sfermion masses. In this case the diagrams listed in
figure \ref{top} are to be inserted into the self-energy radiative corrections to the scalar propagators. Many such contributions should
be computed in this case since now the external lines of the diagrams would not only have gauginos but also gauge bosons and D-fields
of the visible gauge group $G_v$.

In this section we promote $G_v$ to the MSSM gauge group, and consider the
case where the messengers form a complete $SU(5)$ multiplet of index $\ell=1$ for each of the MSSM gauge groups\footnote{For $U(1)_Y$
the ``index'' is customarily defined as $\ell=6Y^2/5$ and the ``casimir" $c_2=3Y^2/5$.}. We still consider $G_h = U(1)$ since more general
cases can be easily accommodated by inserting the appropriate group theory factors. In this case the parameter space spanned by the sfermion
masses is contained in that of minimal gauge
mediation in the sense that all square masses are proportional to one and the same dimension-full parameter. However, the dependence of
this parameter on the dynamics of the hidden sector is quite different from the one arising in minimal gauge mediation and this is the
issue we are going to analyze.

For ease of notation we define, for each type of sfermion, an effective coupling
\be
   g_v^4 = g_1^4 c_2[U(1)] + g_2^4 c_2[SU(2)] + g_3^4 c_2[SU(3)] \label{gvdef}
\ee
in terms of the $SU(3) \times SU(2) \times U(1)$ coupling constants $g_i$ of the Standard Model and the Casimir invariants for each of
the sfermion representations.

A general expression for the mass square was given in \cite{Meade:2008wd}
\be
m_{sf}^2 = - g_v^4 \int \frac{d^4 p }{(2\pi)^4 p^2} \big(C_0^v(p^2/M^2) - 4 C^v_{1/2}(p^2/M^2)
+3 C^v_1(p^2/M^2) \big)~.
\label{ggmorig}
\ee
Separately, each of the functions $C^h_s$ and $C^v_{s^\prime}$ is logarithmically UV-divergent, but all divergences cancel in the final linear combinations. In what follows, we thus only consider the finite parts.
Our first goal is to compute the $C^v_{s^\prime}$ in terms of the $C^h_s$, and this is exactly what the diagrams in figure \ref{top} do.
Obviously there are now two scales ($M$ and $m$) on which $C^v_{s^\prime}$ depends. 
Each of the three functions $C^v_{s^\prime}$ is expressed as linear combination of contributions from the three functions $C^h_s$
\bea
  && C^v_{s^\prime}(p^2/M^2, m^2/M^2)=\int \frac{d^4 l }{(2\pi)^4} \frac{d^4 k }{(2\pi)^4}\big( F_{{s^\prime},0}(l,k,p,m)
   C^h_0(k^2/M^2)- \nn\\
   && 4 F_{{s^\prime},1/2}(l,k,p,m)C^h_{1/2}(k^2/M^2) + 3 F_{{s^\prime},1}(l,k,p,m)C^h_1(k^2/M^2)\big)~, \label{tumb}
\eea
where we have introduced a series of functions $F_{s^\prime,s}(l,k,p,m)$ that can be computed explicitly from the diagrams in Figure~\ref{top}. The four-momentum $k$ always denotes the momentum going through the current-current correlators in the hidden sector.
To obtain the sfermion masses we must then add the three functions $C^v_{s^\prime}$ according to (\ref{ggmorig}) and integrate over the four-momentum $p$. By switching the integrals, leaving the one over $k$ for last, and combining the various functions  $F_{s^\prime,s}(l,k,p,m)$ we can easily rewrite the sfermion masses as
\bea
 m_{sf}^2 &=& \frac{g_v^4 g^4_h}{(4 \pi)^4}\int \frac{d^4 k }{(2\pi)^4 k^2} \big( K_0(k^2/m^2) \, C_0^h(k^2/M^2)- \nn\\
 && 4 K_{1/2}(k^2/m^2) \, C_{1/2}^h(k^2/M^2) + 3 K_1(k^2/m^2) \, C_1^h(k^2/M^2) \big)~, \label{ks}
\eea
where
\bea
     &&  \frac{g^4_h}{(4 \pi)^4 k^2} K_s(k^2/m^2)= \! \nn\\
     &&-\int \frac{d^4 l d^4 p}{(2\pi)^8 p^2} \big( F_{0,s}(l,k,p,m) -
     4 F_{1/2,s}(l,k,p,m) + 3 F_{1,s}(l,k,p,m) \big)~.
\label{kfromf}
\eea
The expressions $K_s(k^2/m^2)$ represent three (a priori independent) scalar ``kernels" depending \emph{only} on the messenger
sector and thus exactly computable irrespectively of what strong dynamics is ultimately responsible for supersymmetry breaking.
Recalling that in the supersymmetric limit all $C^h_s$ are equal and $m^2_{sf}=0$, it follows that the weighted sum of the three kernels
must vanish: $K_0 - 4 K_{1/2} + 3 K_1 = 0$. We have computed explicitly the first two of them and found them to be the same
($K_0 = K_{1/2} \equiv K $), thus implying that the full contribution to the sfermion masses can be written as
\be
\label{mgsd}
m_{sf}^2 =  \frac{g_v^4 \, g^4_h}{(4 \pi)^6} \int d k^2 \, K(k^2/m^2) \, \big(C_0^h(k^2/M^2) - 4 C^h_{1/2}(k^2/M^2) + 3 C^h_1(k^2/M^2) \big)~.
\ee
The kernel $K(x)$ is given by a sum of integrals over two loop momenta, which are exactly known. We relegate to Appendix B the details
of the computation. Suffices here to say that in this case the diagrams of type $b$ involving the messenger four point function are
necessary for consistency, for instance to obtain a transverse vectorial current two point function (i.e. a well-defined $C^v_1$).

\begin{figure}[ht]
\centering
\includegraphics[width=0.85\textwidth]{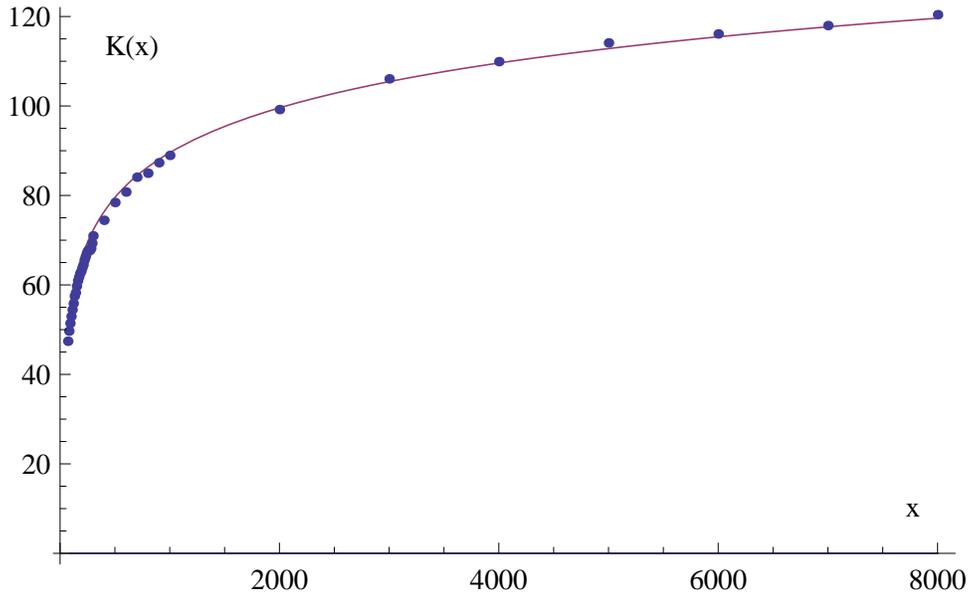}
\caption{\small Numerical estimate of the kernel $K(x)$ for large $x$. The fit is represented by the curve $14.4 \log x - 10.3$.
The numerical errors are about 5\% -- quite sufficient for the simple estimates done in this paper.
\label{kernel}}
\end{figure}

$K(x)$ is a positive function and has the following asymptotic behavior
\bea
K(x) &=& \frac{5}{18} x^2 - \frac{137}{1350} x^3 + \frac{5437}{176400} x^4 + {\mathcal{O}}(x^5)
\quad\mbox{for~~}  x \to 0~, \label{taylorexp} \\ &&\nn  \\
K(x) &\sim&   \gamma\; \log x \quad \mbox{for~~} x \to \infty~, \label{largex}
\eea
where we have estimated numerically $\gamma \sim 14.4$. A plot of the kernel at large $x$ is given in figure \ref{kernel}.

From the behavior of the kernel it is clear that the $m^2_{sf}$ will always be finite since the weighted sum of the $C^h_s$ is soft
enough at large momenta, as noted in \cite{Meade:2008wd}. Note however some important facts.

First of all, in our set up we have a non vanishing supertrace in the messenger sector. According to the argument of
\cite{Poppitz:1996xw} this fact, by itself, leads to an enhancement of sfermion squared masses proportional to the logarithm of a UV scale.
Here we do not see any such dangerous enhancement. Since the theory is renormalizable and no soft terms appear in the bare Lagrangian,
there cannot be counterterms for the sfermion masses and the only two scales of the problem are $M$, the hidden
sector supersymmetry breaking scale, and $m$, the messenger supersymmetric mass.

Secondly, due to the fact that $K(x)$ is positive,
the sfermion squared masses will be of the opposite  sign
with respect to the case where the same hidden supersymmetric sector
is directly coupled to the MSSM, cfr. eqs.~(\ref{ggmorig}) and (\ref{mgsd}). As we will see in the
next section, this is something useful when trying to use SDGM in
concrete phenomenological models.

As a last remark, we note that the fact the kernel vanishes for
small momenta leads to a vanishing contribution to the sfermion masses
from $D$-terms, if such terms are present. One would expect a first contribution from
them at order $(D^h)^2$, since the contribution at linear order
vanishes after summing over both charge conjugate messengers. Two
$D$-tadpole insertions can actually be encoded in
$C^h_{0D}(k^2) \sim (D^h)^2 \delta^4 (k)$ (indeed, two $D$-tadpoles are
equivalent to a $D$-line cut in two, and hence with no momentum flowing
through it). Obviously, this
contribution vanishes when multiplied by $K(x)\sim x^2$. This fact can
be  extracted from the general expression given in
\cite{Poppitz:1996xw} by tuning the parameters so that only $D$-terms
contribute to the messenger mass matrix. This precise situation and
its phenomenology was
analyzed more recently for instance
in \cite{Nakayama:2007cf} where next-to-leading order corrections
were discussed.

We can give a rough estimate of the scaling of $m^2_{sf}$ as a
function of $M$ and $m$, in the two opposite hierarchical limits. We
will assume that the weighted sum is a simple step function (which
can mimic, roughly, the soft behavior previously discussed)
\be
\label{gabri0}
(C^h_0-4C^h_{1/2}+3C^h_1)(k^2/M^2) \sim \frac{1}{16 \pi^2}\Theta(1-k^2/M^2)~,
\ee
with a proportionality factor of either
sign depending on the specific hidden sector\footnote{Again, for the record, we note that in a minimal gauge
mediation scenario we  would obtain $(C_0-4C_{1/2}+3C_1)(p^2) =
(1/16\pi^2)(2 |F|^2/m^4)f'(p^2/m^2)$, where $f'(x)$ is the derivative
of the function $f(x)$ that we encountered previously, and is
everywhere negative.}.

In the limit of heavy messengers we get
\be m^2_{sf} \sim
\frac{\alpha_v^2 \alpha_h^2}{(4 \pi)^4}
\frac{5}{54}\frac{M^6}{m^4}\qquad \qquad m \gg M~, \label{heavymess}
\ee
while in the opposite limit of light messengers
\be
\label{gabri}
m^2_{sf} \sim \frac{\alpha_v^2
\alpha_h^2}{(4 \pi)^4} \gamma M^2 \log \big( \frac{M^2}{m^2}\big)
\qquad\qquad m \ll M~,
\ee
where we have defined $\alpha_{h,v} =
g^2_{h,v}/4\pi$. While we see a signal of the log-enhancement of
\cite{Poppitz:1996xw} in the limit $m \ll M$, we do  not see anything
like it in the opposite limit.

For completeness, we can estimate the
contributions to the mass matrix of the messengers in the limit $m \ll M$ where the radiative corrections can be comparable to the
supersymmetric mass. In this limit we can treat $m$ as a small
perturbation and write the following expressions
\bea
m^2_d & = & - g_h^4 \int \frac{d^4 k }{(2\pi)^4 k^2} \big(C_0^h(k^2/M^2) -
4 C^h_{1/2}(k^2/M^2) + 3 C^h_1(k^2/M^2) \big)~, \\
m^2_o & = & g_h^4 \int \frac{d^4 k }{(2\pi)^4 k^2}
 \frac{m M B^h(k^2/M^2)}{k^2}~,
\eea
where $m^2_d$ and $m^2_o$ are respectively the contributions to the
diagonal and off-diagonal elements of the messenger mass matrix. The supertrace over the messenger sector is
proportional to  $m^2_d$.

Using the estimate (\ref{gabri0}) and a similar one for $B^h$, we obtain
the estimates for $m\ll M$
\be
\label{riccardo}
m^2_d \sim \frac{\alpha_h^2}{(4 \pi)^2} M^2~, \qquad
m^2_o \sim \frac{\alpha_h^2}{(4 \pi)^2} m M \log \big( \frac{M^2}{m^2}\big)~.
\ee
The sign of $m^2_d$ (and thus of the supertrace) is  the opposite of the
sign in the sfermion masses (\ref{gabri}).

We conclude that even if the scale  of the supersymmetry breaking sector is
much higher than the supersymmetric mass of the messengers,  it is
generally possible to avoid tachyonic eigenvalues in their mass matrix
due to the suppression in $\alpha_h$ of the radiative  corrections to
the messenger mass.

\section{Phenomenological uses of semi-direct gauge mediation}

We now comment on the possible phenomenological relevance of the class
of models considered in this paper. We have seen that the sparticle
spectrum produced is essentially one of a strong
hierarchy between the sfermions, which are heavy, and the gauginos which
are very light (massless at the order considered above). This is clearly
not a satisfying spectrum in itself. Additionally, since there is an extra
loop factor $\alpha_h$ in the expression for the
sfermion masses, the gravitino is going to be a factor of $1/\alpha_h$ heavier
than in an ordinary gauge mediation model yielding the same
values of $m_{sf}$.

However, since SDGM gives a contribution mainly to sfermion masses,
it can be useful if combined with other
supersymmetry breaking mechanisms which
provide gaugino masses but tachyonic sfermions.
We review below two such situations, in which the other mechanisms are respectively
anomaly mediation (AM) and direct gauge mediation (DGM).

We first analyze the scenario
where SDGM could address the negative squared sfermion mass problem
of anomaly mediation~\cite{Randall:1998uk}.
We consider the simple set up where
AM and SDGM have the same supersymmetry breaking sector,
and hence the same supersymmetry breaking scale $M$.
Anomaly mediation gives a gaugino mass of the order
\be
m_{\lambda} \sim \frac{\alpha_v}{4 \pi} \frac{M^2}{M_{pl}}~. \label{gauginomasses}
\ee
Assuming a sequestered hidden sector, the slepton masses for the first two generations are of the order
\footnote{Since we are only interested in order of magnitude estimates,
we use a notation similar to (\ref{gvdef}) and lump all dependence on the visible couplings into
$\frac{\alpha_v}{4 \pi}\sim 3 \cdot 10^{-3}$.
One could be more precise, if needed. For instance, focusing on $U(1)_Y$, one has, for the Bino and the right selectron:
$m_{\tilde B} = (33/5) \frac{\alpha_1}{4 \pi} \frac{M^2}{M_{pl}}$ and
$m^2_{{\tilde e}_R} = - (198/25)\frac{\alpha_1^2}{(4 \pi)^2}
\frac{M^4}{M_{pl}^2}$ respectively (see e.g.~\cite{Martin:1997ns}).}
\be
\label{AMmasse}
m_{AM}^2 \sim - \frac{\alpha_v^2}{(4 \pi)^2} \frac{M^4}{M_{pl}^2}~.
\ee
The slepton masses in (\ref{AMmasse}) are tachyonic because
of the sign of the beta function coefficient (the contribution from the Yukawa couplings can be ignored for the
first two generations).

We can cure the problem of tachyonic sleptons by combining AM with a
SDGM model which yields positive sfermion squared masses.
Let us denote by $\delta$ the ratio between the SDGM
contribution (\ref{heavymess}) or (\ref{gabri}) to the slepton masses and (\ref{AMmasse}). Depending on the messenger
scale, we obtain
\be
\delta \equiv \left|\frac{m_{SD}^2}{m_{AM}^2}\right| \sim \left\{\begin{array}{cc}
\frac{\alpha_h^2}{(4 \pi)^2} \gamma \frac{M_{pl}^2}{M^2} \log \frac{M^2}{m^2}
& \hbox{for~~} m \ll M\\ \\
 \frac{\alpha_h^2}{(4 \pi)^2}\frac{5}{54}\frac{M^2 M_{pl}^2}{m^4} & \hbox{for~~} m \gg M \end{array} \right.
\ee
We require $1<\delta<10$ in order for the SDGM
to give a contribution to sfermion masses which is larger than AM but
of the same order.
This can be achieved in both regimes while staying at weak coupling in $\alpha_h$.

First of all, in order to have gaugino masses (\ref{gauginomasses}) at the~TeV scale, we need to take $M \sim 10^{12}$~GeV.

In the first case ($m \ll M$), assuming $\frac{\alpha_h}{4\pi} \sim 10^{-7}$
gives a sensible soft spectrum where the messenger mass can be anywhere in the range $10^5-10^{10}$~GeV. Note that such
a small $G_h$ coupling constant could be actually related to the mechanism of sequestering itself.
In the second case ($m \gg M$), one could have for instance $m \sim 10^{14}$~GeV with $\frac{\alpha_h}{4 \pi}\sim 10^{-2}$.

We conclude that SDGM can be successfully combined with AM, in both regimes and with no substantial fine-tuning, and cure the
slepton problem (some earlier interesting proposals to cure this problem can be found in \cite{Pomarol:1999ie}, and more
recently in \cite{deBlas:2009vx}). It is worth mentioning at this point that such a conspiracy could arise naturally in
string theory since SDGM is generic in D-branes embeddings of gauge mediation \cite{Argurio:2009pz}, while anomaly mediation
must always be included once gravity effects are considered on D-branes.

We now turn to the study of the combination of SDGM with DGM.
We consider again the simple set up where
the two mechanisms share the same
supersymmetry breaking sector.
Note that the currents coupling to the
visible gauge group $G_v$ and the ones
coupling to $G_h$ must be different since they couple to different gauge groups. However we work in the approximation where
their correlation functions are essentially the same. This is not unnatural if the supersymmetry breaking sector has only
one scale or if the two groups arise from the breaking of a larger group.
The important point here is that the contribution to the sfermion masses of SDGM
has an opposite sign with respect to the DGM one. This comes, as already
noticed in section~4, by comparing eqs.~(\ref{ggmorig}) and (\ref{mgsd}), given the positivity
of the kernel $K(x)$.

Models of strongly coupled DGM can lead to unsuppressed gaugino masses
but negative squared scalar masses.
Using the approximation (\ref{gabri0}) the DGM contribution to the sfermion masses is
\be
m_D^2 \sim - \frac{\alpha_v^2}{(4 \pi)^2} M^2~. \label{md2}
\ee
In this case the positive SDGM contribution could render the
sfermion masses non tachyonic, in the limit $m \ll M $ where it can be larger than (\ref{md2}).
To understand if the competition between
SDGM and DGM can be realized naturally, we need to estimate the ratio of the sfermion mass contributions (\ref{gabri}) and (\ref{md2})
\be
\label{delta_1}
\delta=\left|\frac{m_{SD}^2}{m_D^2}\right| \sim \gamma \frac{\alpha_h^2}{(4\pi)^2} \log \frac{M^2}{m^2}~.
\ee
We would then demand $1< \delta  <10$, as before.

Note that, besides the requirement on $\delta$, we have to check that the messengers in the SDGM sector
do not become tachyonic due to radiative corrections.
In section~4 we estimated the diagonal and off-diagonal corrections to the messengers
mass matrix (\ref{riccardo}) in the limit $m \ll M$.
Note that in the case at hand the diagonal radiative correction $m_d^2$
is negative since the SDGM contribution to sfermion masses is positive.

It is possible, in principle, to satisfy these competing constraints. However, some amount of tuning will be
needed in this case. On the one hand, the scale $M$ should be larger than $m$ to avoid a too strong
$G_h$ coupling while keeping $\delta > 1$. On the other hand, $M$ should not be too large, in order to avoid tachyons in the messenger sector.
The possibility of satisfying both these constraints is not generic and it can only be answered on a case by case basis. However, the possibility
of a window where such a mechanism can work is not ruled out.\footnote{Possibly, combining SDGM and DGM with
different supersymmetry breaking sectors can ameliorate
these problems, at the price of introducing new scales in the model.}

An alternative scenario concerns models of DGM which
present an MSSM sparticle spectrum where the gaugino mass
is suppressed or of the same order of the (positive) sfermion masses.
Here the SDGM can provide a negative contribution
to the sfermion masses in order to invert this hierarchy.
This scenario can be realized similarly as above, however with a fine tuning of $\delta$,
i.e.~of the coupling
constant $\alpha_h$.
The fact that a fine tuning is needed in order to obtain
gauginos more massive than sfermions
seems a common feature of gauge mediated models (see for instance \cite{Buican:2008ws}).

In conclusion, our preliminary analysis indicates that models of AM+SDGM can
naturally lead to a sensible MSSM soft mass spectrum and thus seem promising for phenomenological
applications. On the other hand, the SDGM+DGM scenarios that we discussed above
can possibly lead to sensible phenomenology only in a small region of the parameter space, if at all.

\subsection*{Acknowledgments}
We thank Ken Intriligator for discussions at the beginning of this project and Andrea Romanino for useful comments
on a preliminary version of the draft. The research of R.A. is supported in part by IISN-Belgium (conventions
4.4511.06, 4.4505.86 and 4.4514.08). R.A. is a
Research Associate of the Fonds
de la Recherche Scientifique--F.N.R.S. (Belgium).
The research of G.F. is supported in part by the Swedish Research Council (Vetenskapsr{\aa}det)
contract 621-2006-3337. Contribution from the L\"angmanska Kulturfonden and the Wilhelm och Martina Lundgrens
Veteskapsfond are also gratefully acknowledged.
A.M. is a Postdoctoral Researcher of FWO-Vlaanderen.
A.M. is also supported in part by
FWO-Vlaanderen through project G.0428.06.
R.A. and A.M. are supported in part
by the Belgian Federal Science Policy Office
through the Interuniversity
Attraction Pole IAP VI/11.

\appendix
\section{Conventions}
We use the following propagators for the messengers
\bea
\langle \phi (p) \phi^*(-p) \rangle & = & \frac{1}{p^2+m^2} \\
\langle \psi_\alpha (p) \bar \psi_{\dot\alpha}(-p) \rangle & = & \frac{p_\mu \sigma^\mu_{\alpha\dot\alpha}}{p^2+m^2} \\
\langle \psi_\alpha(p) \psi^\beta (-p)\rangle & = & \frac{m\delta_\alpha^\beta}{p^2+m^2}~,
\eea
while for the hidden gauge sector we have the following propagators, at first order in the insertions of the
(supersymmetry breaking) currents
\bea
\langle D^h (p) D^h(-p) \rangle & = & C^h_0(p^2/M^2) \\
 \langle \lambda_\alpha^h (p) \bar \lambda^h_{\dot\alpha}(-p) \rangle & = &- \frac{p_\mu \sigma^\mu_{\alpha\dot\alpha}}
{p^2} C^h_{1/2}(p^2/M^2) \\
  \langle \lambda_\alpha^h (p) \lambda^{h\beta} (-p) \rangle &= & \frac{M \delta_\alpha^\beta}{p^2} B^{h}(p^2/M^2)^\star \\
  \langle A^h_\mu (p) A^h_\nu (-p) \rangle & = & \frac{p_\mu p_\nu - p^2 \eta_{\mu\nu}}{p^4} C^h_1(p^2/M^2)~.
\eea
In these conventions that use exclusively Weyl spinors, each $\phi^*\psi\lambda$ Yukawa vertex comes with a $g\sqrt{2}$
coupling. Note also that
\be
\sigma^\mu_{\alpha \dot \alpha} \bar \sigma^{\nu \dot\alpha \beta} +
\sigma^\nu_{\alpha \dot \alpha} \bar \sigma^{\mu \dot\alpha \beta} = - 2 \eta^{\mu\nu}\delta_\alpha^\beta~,
\ee
and similarly for $\bar \sigma \sigma$.  We use the $(-+++)$
signature, hence all the above relations
are unchanged after Wick rotation.

\section{Computation of the kernel}

In this appendix we collect some more details concerning the computation of the kernels $K_s(x)$, proving, eventually, that
$K_0(x)=K_{1/2}(x)=K_1(x)$. Recall that upon use of the tumbling equations (\ref{tumb}), the sfermion masses (\ref{ggmorig})
depend on the kernels $K_s$ as given in (\ref{ks}).
To compute the kernels $K_s$ one
should write down all the graphs contributing to the visible $C^v_{s'}$ as
in the tumbling
equations (\ref{tumb}). Then one can extract the functions $F_{s',s}$ and
integrate them in order to obtain the kernels (\ref{kfromf}).

\begin{figure}[ht]
\centering
\hspace{1cm}
\includegraphics[width=1\textwidth]{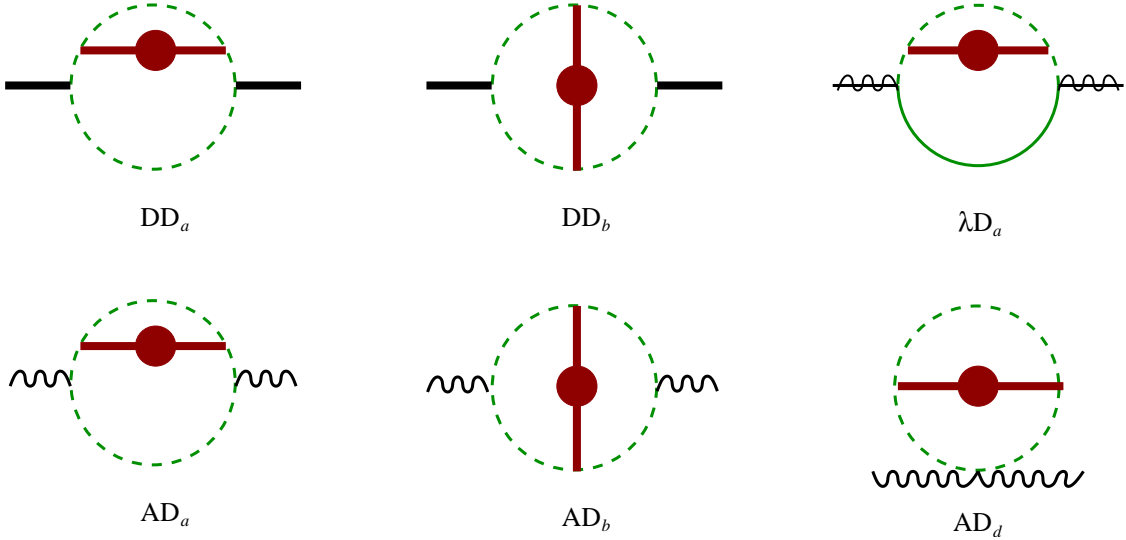}
\caption{\small The diagrams contributing to $K_0$.  The external
lines represent the auxiliary $D$-field, the gaugino and the gauge boson of the visible sector gauge group $G_v$. The
scalar and spinor messengers (dashed and continuum thin lines, respectively) circulate in the outer loop and the bubble
represents the insertion of $C_0^h$. Each diagram must be counted with the appropriate coefficient representing the
different ways messengers can be inserted.
\label{K0}}
\end{figure}

Let us focus on the contributions to $K_0$, first. Since this is defined as the function multiplying $C_0^h$ we need to consider all the diagrams giving the dependence of $C_s^v$ on $C_0^h$, namely those diagrams depicted in figure~\ref{K0}.

\vskip 10pt

The incoming lines represent the visible gauge particles ($D$-field, gaugino and gauge boson),
the particles
going around the external loop are the bosonic and fermionic messengers and, finally, the internal line represents the insertion
of the hidden two-point function $\langle J^h(k) J^h(-k)\rangle$ on a
hidden $D$-line.

To completely specify a diagram one would also need to specify the orientation of the internal lines (including the chirality
type for the fermions) and the type of messengers ($\Phi$ or $\tilde\Phi$). We choose not to write this explicitly in order to
keep the notation simple. Each of the six diagrams in figure~\ref{K0} thus represents a set of diagrams and this is
reflected into the numerical coefficients for each contribution. Other numerical coefficients arise from the normalization of
the interaction vertices and from the Dirac algebra.

The total contribution for each class of diagrams is given by
\bea
\label{diagk0}
DD_a &=&4~ \int  \frac{dk^4 dl^4}{(2\pi)^8} ~\frac{C_0^h(k^2/M^2)}{(l^2+m^2)^2 \, [(l-k)^2 + m^2]\,[(l-p)^2 + m^2]}  \nn \\
DD_b &=&2~ \int  \frac{dk^4 dl^4}{(2\pi)^8} ~\frac{C_0^h(k^2/M^2)}{(l^2+m^2)\,[(l-k)^2 + m^2]\,[(l-k-p)^2 + m^2]\,[(l-p)^2 + m^2]}
\non \\  \non \\
\lambda D_a &=& - 4~\int  \frac{dk^4 dl^4}{(2\pi)^8} ~ \frac{(l-p)_\mu \,\sigma^\mu_{\alpha \dot\alpha} ~ C_0^h(k^2/M^2)}
{(l^2+m^2)^2 \, [(l-k)^2 + m^2]\,[(l-p)^2 + m^2]}
 \nn \\
A D_a &=& 4~\int  \frac{dk^4 dl^4}{(2\pi)^8} ~\frac{(2l-p) _\mu (2l-p)_\nu ~ C_0^h(k^2/M^2)}{(l^2+m^2)^2 [(l-k)^2 + m^2]\,[(l-p)^2 + m^2]}
\label{diag0}\\
A D_b &=& 2~\int  \frac{dk^4 dl^4}{(2\pi)^8} ~\frac{(2l-p) _\mu (2l-2k-p)_\nu ~ C_0^h(k^2/M^2)}
{(l^2+m^2)~ [(l-k)^2 + m^2]\,[(l-k-p)^2 + m^2]\,[(l-p)^2 + m^2]} \nn \\
A D_d &=& - 4~ \int  \frac{dk^4 dl^4}{(2\pi)^8} ~\frac{\eta_{\mu\nu} ~ C_0^h(k^2/M^2)}{(l^2+m^2)^2 \,[(l-k)^2 + m^2]} ~. \nn
\eea
The notation of the above integrals should be self-evident, for instance $AD$ denotes the contribution of the hidden $D$-field to
the visible gauge boson $A$. The subscript refers to
the five topologies introduced in figure~\ref{top} (only diagrams of topology $a$, $b$ and $d$ enter the computation of $K_0$). There are
several consistency checks for the above expressions. For instance, one can check that the sum $A D_a + A D_b + A D_d$ obeys the Ward identities.

To obtain $K_0$ we need to extract the expression for the contribution to $C_s^v$ from the diagrams above by
comparing (\ref{diag0}) with the definition (\ref{firstequation}). One then inserts the contribution thus obtained into
(\ref{ggmorig}). The resulting two-loop integral (in $p$ and $l$) defines the kernel $K_0(k^2/m^2)$. It
turns out to be convenient to write everything with a common denominator and express the integral over $l$
in terms of three Feynman parameters. The resulting integrand can be expanded in power series of $k$ and
the coefficients (depending on the three Feynman parameters, $p$ and the angle between $p$ and $k$) can be
fully integrated yielding (\ref{taylorexp}). Alternatively, one can perform a numerical integration and
obtain the behavior (\ref{largex}) of the kernel for large $k$.

We performed an analogous computation for $K_{1/2}$. (We spare the reader the details.) Despite
the rather different structure of the  contributions, we find that $K_{1/2}$ is the same as $K_0$. As discussed in section 4,
supersymmetry implies that the last kernel $K_1$ also be the same as the previous two thus completing the computation and proving that
we have one and only one common kernel $K$.



\begin{thebibliography}{99}

\bibitem{original}
  M.~Dine and W.~Fischler,
  Phys.\ Lett.\  B {\bf 110} (1982) 227.

  C.~R.~Nappi and B.~A.~Ovrut,
  Phys.\ Lett.\  B {\bf 113}, 175 (1982).

  M.~Dine and W.~Fischler,
  Nucl.\ Phys.\  B {\bf 204} (1982) 346.

  L.~Alvarez-Gaume, M.~Claudson and M.~B.~Wise,
  Nucl.\ Phys.\  B {\bf 207}, 96 (1982).

\bibitem{Giudice:1998bp}
  G.~F.~Giudice and R.~Rattazzi,
  Phys.\ Rept.\  {\bf 322} (1999) 419
  [arXiv:hep-ph/9801271].

\bibitem{Meade:2008wd}
  P.~Meade, N.~Seiberg and D.~Shih,
  arXiv:0801.3278 [hep-ph].

\bibitem{Buican:2008ws}
  M.~Buican, P.~Meade, N.~Seiberg and D.~Shih,
  JHEP {\bf 0903} (2009) 016
  [arXiv:0812.3668 [hep-ph]].

  D.~Marques,
  JHEP {\bf 0903} (2009) 038
  [arXiv:0901.1326 [hep-ph]].

\bibitem{Abel:2009ze}
  S.~A.~Abel, J.~Jaeckel and V.~V.~Khoze,
  arXiv:0907.0658 [hep-ph].

  S.~Abel, M.~J.~Dolan, J.~Jaeckel and V.~V.~Khoze,
  arXiv:0910.2674 [hep-ph].

\bibitem{messengers}
  M.~Dine and A.~E.~Nelson,
  Phys.\ Rev.\  D {\bf 48}, 1277 (1993)
  [arXiv:hep-ph/9303230].

  M.~Dine, A.~E.~Nelson and Y.~Shirman,
  Phys.\ Rev.\  D {\bf 51}, 1362 (1995)
  [arXiv:hep-ph/9408384].

  M.~Dine, A.~E.~Nelson, Y.~Nir and Y.~Shirman,
  Phys.\ Rev.\  D {\bf 53}, 2658 (1996)
  [arXiv:hep-ph/9507378].


\bibitem{direct}
  E.~Poppitz and S.~P.~Trivedi,
  Phys.\ Rev.\  D {\bf 55}, 5508 (1997)
  [arXiv:hep-ph/9609529].

  N.~Arkani-Hamed, J.~March-Russell and H.~Murayama,
  Nucl.\ Phys.\  B {\bf 509}, 3 (1998)
  [arXiv:hep-ph/9701286].

  H.~Murayama,
  Phys.\ Rev.\ Lett.\  {\bf 79}, 18 (1997)
  [arXiv:hep-ph/9705271].

  S.~Dimopoulos, G.~R.~Dvali, R.~Rattazzi and G.~F.~Giudice,
  Nucl.\ Phys.\  B {\bf 510} (1998) 12
  [arXiv:hep-ph/9705307].

\bibitem{koolike}
  R.~Kitano, H.~Ooguri and Y.~Ookouchi,
  Rev.\  D {\bf 75} (2007) 045022
  [arXiv:hep-ph/0612139].

  C.~Csaki, Y.~Shirman and J.~Terning,
  JHEP {\bf 0705} (2007) 099
  [arXiv:hep-ph/0612241].

  A.~Amariti, L.~Girardello and A.~Mariotti,
  Fortsch.\ Phys.\  {\bf 55} (2007) 627
  [arXiv:hep-th/0701121].

  N.~Haba and N.~Maru,
  Phys.\ Rev.\  D {\bf 76} (2007) 115019
  [arXiv:0709.2945 [hep-ph]].

\bibitem{Seiberg:2008qj}
  N.~Seiberg, T.~Volansky and B.~Wecht,
  [arXiv:0809.4437 [hep-ph]].

  H.~Elvang and B.~Wecht,
  JHEP {\bf 0906} (2009) 026
  [arXiv:0904.4431 [hep-ph]].

\bibitem{Randall:1996zi}
  L.~Randall,
  Nucl.\ Phys.\  B {\bf 495} (1997) 37
  [arXiv:hep-ph/9612426].

  C.~Csaki, L.~Randall and W.~Skiba,
  Phys.\ Rev.\  D {\bf 57} (1998) 383
  [arXiv:hep-ph/9707386].


\bibitem{Argurio:2009pz}
  R.~Argurio, M.~Bertolini, G.~Ferretti and A.~Mariotti,
  Phys.\ Rev.\  D {\bf 80} (2009) 045001
  [arXiv:0906.0727 [hep-th]].

\bibitem{ArkaniHamed:1998kj}
  N.~Arkani-Hamed, G.~F.~Giudice, M.~A.~Luty and R.~Rattazzi,
  Phys.\ Rev.\  D {\bf 58} (1998) 115005
  [arXiv:hep-ph/9803290].

\bibitem{Ibe:2009bh}
  M.~Ibe, K.~I.~Izawa and Y.~Nakai,
  arXiv:0812.4089 [hep-ph].

  Z.~Komargodski and D.~Shih,
  JHEP {\bf 0904}, 093 (2009)
  [arXiv:0902.0030 [hep-th]].

  M.~Ibe, K.~I.~Izawa and Y.~Nakai,
  Phys.\ Rev.\  D {\bf 80} (2009) 035002
  [arXiv:0907.2970 [hep-ph]].

  A.~Giveon, A.~Katz and Z.~Komargodski,
  JHEP {\bf 0907}, 099 (2009)
  [arXiv:0905.3387 [hep-th]].


\bibitem{Poppitz:1996xw}
  E.~Poppitz and S.~P.~Trivedi,
  Phys.\ Lett.\  B {\bf 401} (1997) 38
  [arXiv:hep-ph/9703246].

\bibitem{Randall:1998uk}
  L.~Randall and R.~Sundrum,
  Nucl.\ Phys.\  B {\bf 557} (1999) 79
  [arXiv:hep-th/9810155].

  G.~F.~Giudice, M.~A.~Luty, H.~Murayama and R.~Rattazzi,
  JHEP {\bf 9812} (1998) 027
  [arXiv:hep-ph/9810442].

\bibitem{Ibe:2007wp}
  M.~Ibe, Y.~Nakayama and T.~T.~Yanagida,
  Phys.\ Lett.\  B {\bf 649} (2007) 292
  [arXiv:hep-ph/0703110].

  M.~Ibe, Y.~Nakayama and T.~T.~Yanagida,
  Phys.\ Lett.\  B {\bf 671} (2009) 378
  [arXiv:0804.0636 [hep-ph]].

\bibitem{Buican:2009vv}
  M.~Buican and Z.~Komargodski,
  arXiv:0909.4824 [hep-ph].

\bibitem{Nakayama:2007cf}
  Y.~Nakayama, M.~Taki, T.~Watari and T.~T.~Yanagida,
  Phys.\ Lett.\  B {\bf 655} (2007) 58
  [arXiv:0705.0865 [hep-ph]].

\bibitem{Martin:1997ns}
  S.~P.~Martin,
  arXiv:hep-ph/9709356.


\bibitem{Pomarol:1999ie}
  A.~Pomarol and R.~Rattazzi,
  JHEP {\bf 9905} (1999) 013
  [arXiv:hep-ph/9903448].

  Z.~Chacko, M.~A.~Luty, I.~Maksymyk and E.~Ponton,
  JHEP {\bf 0004} (2000) 001
  [arXiv:hep-ph/9905390].

  D.~E.~Kaplan and G.~D.~Kribs,
  JHEP {\bf 0009} (2000) 048
  [arXiv:hep-ph/0009195].


\bibitem{deBlas:2009vx}
  J.~de Blas, P.~Langacker, G.~Paz and L.~T.~Wang,
  arXiv:0911.1996 [hep-ph].

\end{thebibliography}
\end{document}